\documentclass[conference]{IEEEtran}
\IEEEoverridecommandlockouts
\usepackage{cite}
\usepackage{amsmath,amssymb,amsfonts}
\usepackage{algorithmic}
\usepackage{graphicx}
\usepackage{textcomp}
\usepackage{xcolor}
\usepackage{booktabs}
\def\BibTeX{{\rm B\kern-.05em{\sc i\kern-.025em b}\kern-.08em
		T\kern-.1667em\lower.7ex\hbox{E}\kern-.125emX}}
\begin{document}
	\bstctlcite{IEEEexample:BSTcontrol}
	
	\title{Reconstruction of 3-Axis Seismocardiogram from Right-to-Left and Head-to-Foot Components Using A Long Short-Term Memory Network
		\thanks{*This work was supported by Mississippi State University's Office of Research and Economic Development through the Quick Grants Program.}
	}
	
	\author{\IEEEauthorblockN{Mohammad Muntasir Rahman}
		\IEEEauthorblockA{\textit{Dept of Ag. \& Biological Engineering} \\
			\textit{Mississippi State University}\\
			Mississippi State, MS 39762, USA \\
			mmr510@msstate.edu}
		\and
		\IEEEauthorblockN{Amirtahà Taebi}\thanks{Corresponding author: ataebi@abe.msstate.edu}
		\IEEEauthorblockA{\textit{Dept of Ag. \& Biological Engineering} \\
			\textit{Mississippi State University}\\
			Mississippi State, MS 39762, USA \\
			ataebi@abe.msstate.edu}		
	}
	
	\maketitle
	
	\begin{abstract}
		This pilot study aims to develop a deep learning model for predicting seismocardiograms (SCGs) in the dorsoventral direction from the SCG signals in the right-to-left and head-to-foot directions (SCG\textsubscript{x} and SCG\textsubscript{y}). The dataset used for the training and validation of the model was obtained from 15 healthy adult subjects. The SCG signals were recorded using tri-axial accelerometers placed on the chest of each subject. The signals were then segmented using electrocardiogram R waves, and the segments were downsampled, normalized, and centered around zero. The resulting dataset was used to train and validate a long short-term memory (LSTM) network with two layers and a dropout layer to prevent overfitting. The network took as input 100-time steps of SCG\textsubscript{x} and SCG\textsubscript{y}, representing one cardiac cycle, and outputted a vector that mapped to the target variable being predicted. The results showed that the LSTM model had a mean square error of 0.09 between the predicted and actual SCG segments in the dorsoventral direction. The study demonstrates the potential of deep learning models for reconstructing 3-axis SCG signals using the data obtained from dual-axis accelerometers.
	\end{abstract}
	
	
	\section{Introduction}
	Cardiovascular diseases (CVDs) are the leading cause of death in the United States, claiming the life of one person every 34 seconds, resulting in a staggering 2,544 deaths per day based on 2020 data \cite{tsao2023heart}. Beyond the devastating human toll, this places an immense strain on healthcare systems and society, with significant economic costs associated with treatment and lost productivity. The early detection of cardiac abnormalities is crucial for achieving better outcomes for patients with CVD, and improving diagnostic methods and accessibility is a key step in this direction. Current diagnostic methods, including non-invasive techniques such as electrocardiography (ECG), medical imaging, and cardiac catheterization, can aid in identifying CVDs. Advancements in technology have also led to the development of new diagnostic options, such as wearable and remote monitoring systems, which can provide continuous monitoring of patients' cardiovascular health outside of traditional healthcare settings.
	Seismocardiography (SCG) is another technique that noninvasively monitors cardiovascular activity by measuring cardiovascular-induced vibrations on the chest \cite{taebi2017time, zanetti1991seismocardiography,rahman2023non}. These vibrations result from a range of cardiac activities, including valve opening and closing, isovolumetric contraction, blood ejection, and rapid left ventricle filling \cite{zanetti1991seismocardiography,taebi2019recent, mann2024exploring, taebi2017analysis}. Unlike other non-invasive techniques such as ECG and pulse oximetry, which focus on the electrical activity of the heart and the blood oxygen level, respectively, SCG provides complementary insights into the mechanical activity of the heart \cite{inan2014ballistocardiography, taebi2019recent, cook2022body}. With its ability to evaluate these mechanical activities, SCG has the potential to offer valuable diagnostic information for cardiac conditions such as heart failure, myocardial infarction, ischemia, and hemorrhage, as changes in the mechanical function of the heart can be an early indication of these diseases \cite{sorensen2018definition, ha2020contactless, tavakolian2010estimating, pankaala2016detection, shandhi2022estimation, salerno1991seismocardiography}. As a result, SCG can enhance our understanding of the cardiac function and contribute to the development of more accurate diagnostic tools for patients with CVD. SCG signals are commonly measured using accelerometers that are placed on the chest surface. These signals are typically measured in three directions of right-to-left, head-to-foot, and dorsoventral. In that regard, while single or dual-axis accelerometers may be used for SCG measurement, three-axis accelerometers are more informative as they offer a more comprehensive understanding of the motion of the heart and chest wall \cite{inan2014ballistocardiography}. 
	
	This study aims to answer the question: ``Can we generate three-axis SCG measurements using a dual-axis accelerometer?" More specifically, is it possible to generate SCG vibrations in the $z$ direction using the measurements from the $x$ and $y$ axes of a dual-axis accelerometer? In this work, we propose a deep neural network model based on long short-term memory (LSTM) to predict the SCG component in the dorsoventral direction (SCG\textsubscript{z}) using the vibrations in right-to-left and head-to-foot directions (SCG\textsubscript{x}, SCG\textsubscript{y}). LSTM models have been successfully applied to a wide range of time series-related problems, including but not limited to stock price prediction, energy load forecasting, weather forecasting, speech recognition, and natural language processing. In this paper, we designed a regression model based on a stacked LSTM neural network with two layers to process the SCG\textsubscript{x} and SCG\textsubscript{y} sequences corresponding to a single cardiac cycle and generate an output sequence of SCG\textsubscript{z} of the same length. The use of a stacked LSTM network can potentially improve the accuracy of the predictions by allowing the network to learn more complex temporal patterns in the data.

	\section{Materials and Methods}
	
	\begin{figure*}
		\centering
		\includegraphics[width=0.9\textwidth]{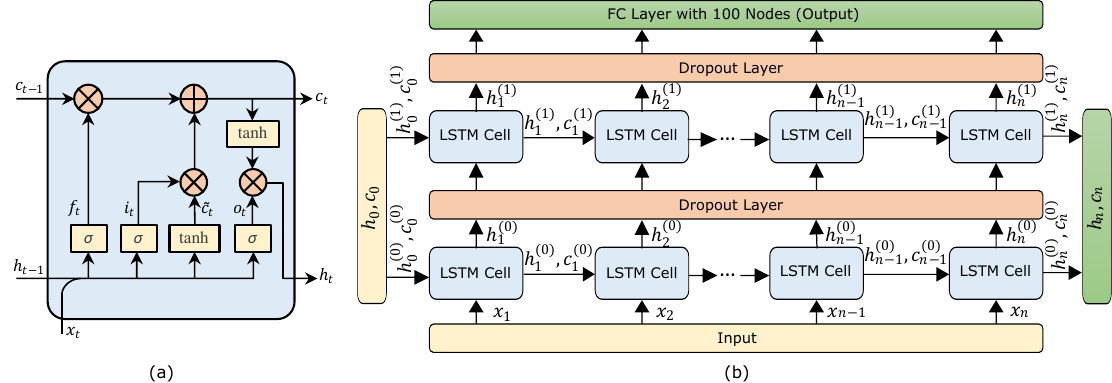}
		\caption{(a) A representation of a LSTM cell.        
			(b) Stacked LSTM network architecture.}
		\label{fig:network}
	\end{figure*}
	
	\subsection{Long Short-Term Memory Network}
	
	The basic building block of an LSTM cell consists of a memory cell and three gates: the input gate, the forget gate, and the output gate. The memory cell is responsible for storing information over time and passing it forward through the sequence. The input gate controls how much new information is added to the cell state, the forget gate controls how much old information is removed from the cell state, and the output gate controls how much information from the cell state is used to compute the hidden state. 
	Fig. \ref{fig:network}a shows a schematic diagram of an LSTM unit. At each time step $t$, the LSTM cell takes as input the current input sequence $x_t$, the previous hidden state $h_{t-1}$, and the previous cell state $c_{t-1}$. Using these inputs, the LSTM cell generates the current hidden state $h_t$ and the updated cell state $c_t$. Let
	\begin{flalign} \qquad
		& i_t = \sigma(W_x^{(i)}x_t + W_h^{(i)}h_{t-1} + b^{(i)})	&\\	
		& f_t = \sigma(W_x^{(f)}x_t + W_h^{(f)}h_{t-1} + b^{(f)})	&\\	
		& o_t = \sigma(W_x^{(o)}x_t + W_h^{(o)}h_{t-1} + b^{(o)})	&\\
		& \tilde{c}_t = tanh(W_x^{(c)}x_t + W_h^{(c)}h_{t-1} + b^{(c)}) &
	\end{flalign}
	where $i_t$, $f_t$, and $o_t$ are the input, forget, and output gates, respectively, and $\tilde{c}_t$ provides the change contents, then the updated cell state $c_t$ and hidden state $h_t$ are computed as: 
	\begin{flalign} \qquad
		& c_t = f_t \odot c_{t-1} + i_t \odot \tilde{c}_t	&\\
		& h_t = o_t \odot tanh(c_t)&
	\end{flalign}
	where $\odot$ performs element-wise product.
	

	\subsection{Study Population}
	The study enrolled a total of 15 participants (including 4 female), with no prior history of cardiovascular diseases (age: 25.93~$\pm$~10.65 year, height: 171.31~$\pm$~9.22 cm, weight: 74.83~$\pm$~22.83 kg, body mass index: 25.29~$\pm$~6.58 kg/m$^2$). The study sample was diverse, with participants from various racial and ethnic backgrounds, including 53.3\% White, 20\% Black, 20\% Asian, and 6.7\% mixed. The Mississippi State University Institutional Review Board approved the study protocol.
	
	\subsection{Data Acquisition Protocol}
	To minimize any potential movement artifacts that could affect the quality of the data, all subjects were instructed to lay supine on a bed without additional body movements. Three triaxial accelerometers (356A32, PCB Piezotronics, Depew, NY) were attached to three locations on the sternum including the manubrium, the fourth costal notch, and the xiphoid process. The accelerometer outputs were amplified using a signal conditioner (482C, PCB Piezotronics, Depew, NY) with a gain factor of 100 to increase the signal-to-noise ratio. The amplified signals were then recorded using a data acquisition system (416, iWorx Systems, Inc., Dover, NH), with a sampling frequency of 5000 Hz. A microphone was connected to the system and tapped at the beginning and end of each recording. These taps were then located in the sound signals to identify the start and end of the intended part of the recording. An ECG module was also used to simultaneously record the ECG signal. 
	Data were collected from all accelerometers during a 15-second breath-hold at the end of inhalation and exhalation, and 2 additional minutes of normal breathing.
	
	
	\subsection{SCG Dataset} \label{sec:dataset}
	To prepare the dataset, the following pre-processing steps were carried out to eliminate noise from the raw signals. The first step involved applying a moving average filter to smooth the SCG signals. This step helps reduce high-frequency noise in the data, making it easier to detect underlying patterns or features. Subsequently, a band-pass filter was applied to the accelerometer outputs with cutoff frequencies of 1 and 30 Hz. This eliminated the low-frequency respiration vibrations and the higher-frequency SCG components above 30 Hz. 
	
	SCG signals were then segmented using the ECG R waves as reference points. To detect the R waves, we utilized Pan-Tompkins algorithm \cite{sedghamiz2014matlab}. We then computed the average duration of the cardiac cycle for each subject from the ECG RR intervals. This information was then used to determine the window size to segment the SCG signals for each subject. Specifically, we set the start of the window to 1/4 of the average cardiac cycle duration before the R wave and the end of the window to 3/4 of the average cardiac cycle duration after the R wave, resulting in consistent SCG segments. After segmenting the SCG signals for each cardiac cycle, the segments were downsampled to a fixed number of sample points (100 points in this study). This allowed us to obtain consistent segment lengths across all samples for different subjects, ensuring that the input data for the LSTM model was uniform. This is crucial because the neural network takes one segment at a time as input, and the original segments may have varied lengths for different subjects (due to different cardiac cycle durations). By treating each SCG segment as a separate sample in the dataset, the neural network can learn patterns and features specific to each cycle, which can be useful for predicting SCG signals in the $z$-direction.
	
	Next, the segments were normalized to have values between -1 and 1 to ensure that the input data falls within a suitable range for activation functions. Then, the mean value was subtracted from each segment to center the signals around zero. This step was useful in mitigating any DC offset or baseline drift present in the signal that could affect the performance of the neural network. In the last step, the pre-processed and normalized SCG segments from all subjects were combined to form a single dataset, which was used for training, validating and testing the neural network model. A total of 7492 SCG segments were used for training and validation and the model was tested on 475 segments. This dataset was expected to capture the common patterns and features present in the SCG signals of the population studied, and was thus representative of the entire cohort.
	
	\subsection{Network Architecture and Training}
	Fig. \ref{fig:network}b provides an overview of a stacked LSTM network's architecture employed in this study. A stacked LSTM refers to the use of multiple LSTM layers, one on top of the other which allows learning more complex patterns and relationships in the data. The network has two stacked LSTM layers, each with a hidden size of 512. The hidden size refers to the number of hidden units or neurons in the LSTM layer, which determines the capacity of the network to capture complex patterns in the data. The network processes 100-time steps of SCG\textsubscript{x} and SCG\textsubscript{y} as a sample input, which corresponds to approximately one cardiac cycle. Utilizing dual-axis rather than single-axis SCG data may help the network to capture more information and learn the relationship between the input (SCG\textsubscript{x} and SCG\textsubscript{y}) to predict the SCG\textsubscript{z} and make accurate predictions on new, unseen data. The input then flows into the stacked LSTM layers. The final layer outputs a vector $h_i$, which is subsequently fed into a fully connected layer featuring 100 output neurons. This fully connected layer maps the input vector to a set of output values that correspond to the target variable predicted by the model. To prevent overfitting and improve the generalization, a dropout layer is added after each LSTM layer. The dataset was divided into training, validation, and testing sets to train and evaluate the network. Since we collected data from three accelerometers for each subject, we separated breath-hold data captured at the end of inhalation and at the end of exhalation from the accelerometer attached to the xiphoid process to test the model. The remaining data were used for training and validation. The training set, which contained 90\% of the data, was used to determine the weight and bias parameters through forward and backward propagation during the training process. The validation set, which contained 10\% of the data, was used to evaluate the model performance during the training process. We trained the network with a maximum of 1000 epochs and used early stopping to prevent overfitting during training. Early stopping is a regularization technique to monitor the performance of the model on the validation set during training and stop the training when the performance on the validation set no longer improves. The initial learning rate was 0.001, the learning rate was reduced using learning rate decay when the validation set stopped improving for a specified number of epochs. This technique helped the model converge to a better solution and avoid getting stuck in local minima.
	

	\begin{table}
		\caption{Accuracy of the model in terms of MSE}
		\label{table_accuracy}	
		\centering
		\begin{tabular}{cccc} 
			\toprule
			\textbf{Subject} & \textbf{Exhale-MSE} & \textbf{Inhale-MSE} & \textbf{Total-MSE} \\
			\midrule
			Subject 01 & 0.064 & 0.083 & 0.07  \\
			Subject 02 & 0.037 & 0.115 & 0.07  \\
			Subject 03 & 0.094 & 0.096 & 0.09  \\
			Subject 04 & 0.069 & 0.071 & 0.07  \\
			Subject 05 & 0.087 & 0.150 & 0.11  \\
			Subject 06 & 0.055 & 0.062 & 0.05  \\
			Subject 07 & 0.107 & 0.057 & 0.08  \\
			Subject 08 & 0.070 & 0.163 & 0.11  \\
			Subject 09 & 0.079 & 0.063 & 0.07  \\
			Subject 10 & 0.073 & 0.087 & 0.08  \\
			Subject 11 & 0.189 & 0.098 & 0.14  \\
			Subject 12 & 0.058 & 0.077 & 0.06  \\
			Subject 13 & 0.129 & 0.107 & 0.11  \\
			Subject 14 & 0.071 & 0.082 & 0.07  \\
			Subject 15 & 0.140 & 0.092 & 0.11  \\
			All Subjects & 0.087 & 0.096 &  0.09 \\
			\bottomrule
		\end{tabular}	
	\end{table}

	\begin{figure*}	
		\centering
		\includegraphics[width=0.9\textwidth]{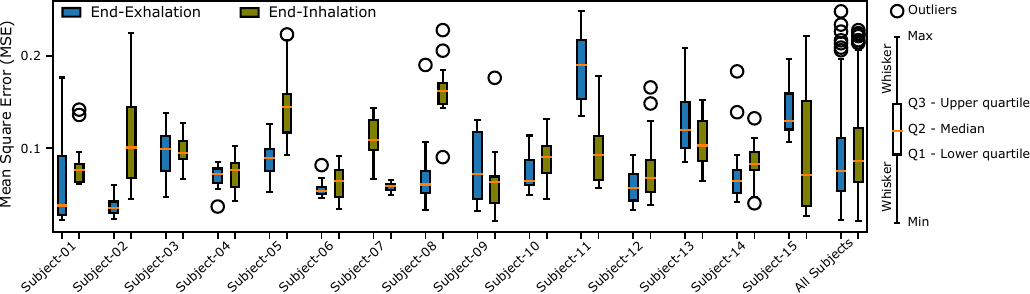}
		\caption{Mean square error between the predicted and actual SCG\textsubscript{z} segments in the test set.}
		\label{fig:box-plot}
	\end{figure*}
	
	\begin{figure*}	
		\centering
		\includegraphics[width=0.95\textwidth]{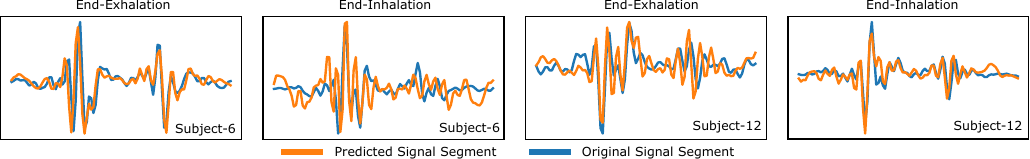}
		\caption{Predicted versus actual SCG\textsubscript{z} segments for Subjects 6 and 12.}
		\label{fig:signal-comparison}
	\end{figure*}
	
	\section{Results and Discussion}
	Our goal was to investigate the possibility of predicting SCG\textsubscript{z} from the SCG signals in $x$ and $y$ directions through the use of an LSTM neural network. We used the testing set to evaluate the model's performance. The accuracy of the model in predicting SCG\textsubscript{z} based on SCG\textsubscript{x} and SCG\textsubscript{y} was assessed by comparing the predicted and actual SCG signal in the $z$ direction, using mean-square error (MSE) as the measure of accuracy. 
	Table \ref{table_accuracy} shows the model's performance in terms of MSE, while Fig. \ref{fig:box-plot} provides a box plot to illustrate the results for each subject's end-exhalation and end-inhalation data, as well as the total performance based on all end-exhalation and end-inhalation data. The box plot shows the median, interquartile range, and outliers of the prediction. The prediction MSE for each subject varied from 0.05 to 0.14, with an MSE of 0.09 for all subjects combined. 
	Four samples of the predicted and actual SCG\textsubscript{z} segments from subjects 6 and 12 are presented in Fig. \ref{fig:signal-comparison}. The close resemblance between the predicted and actual SCG\textsubscript{z} signals suggests that there is a correlation between the SCG components in the $x$, $y$, and $z$ directions, and that our LSTM network was able to learn it.


	\section{Conclusion}
	
	The study investigated the possibility of predicting SCG\textsubscript{z} from SCG components in $x$ and $y$ directions using an LSTM neural network. Results showed that the predicted SCG\textsubscript{z} closely matched the actual signal, indicating a relationship between the SCG components in the three directions. The study provides insights into the potential of using a dual-axis accelerometer to monitor SCG signals in all three directions, which can improve cardiovascular monitoring methods.

	\bibliographystyle{IEEEtran}	
	\bibliography{references}	
\end{document}